\documentclass[12pt]{article}

\usepackage{amssymb,amsmath}

\begin{document}

\begin{flushright}
CU-TP-1071
\\*[25pt]
\end{flushright}

\begin{center}

\textsf{\textbf{\large The Energy Dependence of the Saturation
Momentum
\\*[4pt]
from RG Improved BFKL Evolution}}\footnote {This
research is supported in part by the US Department of Energy}
\\*[20pt]

{D.N.~Triantafyllopoulos}\footnote {E-mail address:
dionysis@phys.columbia.edu}
\\*[2pt]
{\it \small Physics Department, Columbia University,}
\\
{\it \small New York, NY 10027, USA}
\\*[30pt]

\end{center}

\begin{abstract}

We study the energy dependence of the saturation momentum in the
context of the collinearly improved Leading and Next to Leading
BFKL evolution, and in the presence of saturation boundaries. We
find that the logarithmic derivative of the saturation momentum is
varying very slowly with Bjorken-$x$, and its value is in
agreement with the Golec-Biernat and W\"{u}sthoff model in the
relevant $x$ region. The scaling form of the amplitude for
dipole-dipole or dipole-hadron scattering in the perturbative side
of the boundary is given.

\end{abstract}

\section{Introduction}

The idea of parton saturation was introduced twenty years ago from
Gribov, Levin and Ryskin \cite{GLR} as a dual description of
unitarity in high energy hard processes. Since then significant
progress has been made in our understanding of saturation and
unitarity \cite{MIK}. Appropriate QCD evolution equations that
describe high density partonic systems have been derived
\cite{JKLW,Weigert,FILM,Mueller}, and saturation is gradually
becoming well established theoretically. At the same time there
has been considerable phenomenological success, as saturation
based models are used to explain data at RHIC and HERA. The
McLerran-Venugopalan model \cite{McLVen,Kov,JKMW} exhibits gluon
saturation in a simple and intuitive way, and is used in order to
describe the early stages of high energy heavy ion collisions. In
deep inelastic scattering, the model of Golec-Biernat and
W\"{u}sthoff \cite{GBW} has been quite successful, as it gives a
very good fit to the HERA data at small values of $x$ and in a
wide region of $Q^2$ values.

One of the important issues has been the energy (or rapidity)
dependence of the saturation momentum $Q_s(Y)$. How can we
determine $Q_s(Y)$ from the available QCD evolution equations? To
be more specific let's consider the scattering of a color dipole
of size $1/Q$ on a dipole or even a hadron of size $1/\mu$ and
with relative rapidity $Y$. Then, considering the forward
scattering amplitude $T(Q,\mu,Y)$, the saturation momentum
corresponds to a line in the $Q^2-Y$ plane, along which $T$
becomes constant and of order $1/\mu^2$, that is unitarity effects
have been reached.

One should expect BFKL evolution \cite{KLF,BL} (but not its double
logarithmic limit) to be the relevant dynamics, since this is the
one that leads to high density partonic systems. For fixed
coupling and at leading level this was done in \cite{Mueller2},
and the leading exponential (in $Y$) behavior of $Q_s$ was found.
In \cite{IIM} it was reconfirmed and extended to the running
coupling case with the result
$\log(Q_s^2/\Lambda^2)\propto\sqrt{Y}$. BFKL dynamics, being a
linear evolution, suffers from the lack of unitarity. Therefore it
would seem natural, that the use of a more general evolution
equation would give the proper description of $Q_s$. This was done
in \cite{GBMS}, where a numerical study of the Balitsky-Kovchegov
equation \cite{Bal,Kov2} was carried out.

It would be nice of course to have analytical results from such
non-linear equations. However, the exact details of the non-linear
evolution effects should not be important in determining the
energy dependence of $Q_s$. As far as the system remains on the
perturbative side of the saturation line, all the effects of the
non-linear terms could be described by a simple absorptive
boundary. Using leading BFKL with either fixed or running
coupling, this procedure was followed in \cite{MueTri}. In the
more realistic case of running coupling, negative $Y^{1/6}$
corrections to the leading $\sqrt{Y}$ were found. Even though this
correction is parametrically small when compared to the leading
behavior, it becomes important for realistic values of $Y$, and it
tames the growth of $Q_s$ which is overestimated in the absence of
boundaries.

The goal in this paper is the extension of this procedure at the
next to leading level. The calculation of the NL correction to the
BFKL kernel was completed in \cite{FadLip,CamCia}. However, this
negative correction turned out to be larger in magnitude than the
leading contribution for reasonable values of $\alpha_s=0.2\div
0.3$. Even worse, when $\bar{\alpha}_s \gtrsim 0.05$, the full
kernel has two complex saddle points which lead to oscillatory
cross sections \cite{Ross}. Thus there is no hope that one will
get sensible and reliable results when using the standard NL BFKL
dynamics. Perhaps the most appealing cure of these pathologies,
has been given by Ciafaloni, Colferai and Salam
\cite{Salam,CiaCol,CCS,Salam2}. It was recognized that the large
NL corrections emerge from the collinearly enhanced physical
contributions. A method, the $\omega$-expansion, was then
developed to resum collinear effects at all orders, and the
resulting improved BFKL equation was consistent with
renormalization group constraints and the DGLAP equations
\cite{Gribov,Dok,AltPar} by construction. The kernel is positive
in a much larger region which includes the experimentally
accessible one. One of the outcomes of this method was a very
reasonable value for the hard Pomeron intercept $\omega_P$.
Therefore we expect the application of the $\omega$-expansion in
the saturation problem, to give reliable results for the energy
dependence of $Q_s$ and for the form of the scattering amplitude
in the perturbative side of the saturation line.

In Section 2 we give a brief review of the $\omega$-expansion
approach, and the full form of the NL in $\omega$ kernel that we
use is given in the Appendix. In Section 3 we define a critical
line $Q_c(Y)$ (not the saturation line) which belongs in
the perturbative region and which is a line of almost, but not
exactly, constant amplitude; there is no exponential (in $Y$)
dependence. In Section 4, by considering perturbations around this
line, we approximately solve the RG improved BFKL equation under
certain but valid assumptions. In Section 5 we introduce the
boundary, which essentially cuts out all paths that enter the
saturation region $Q \lesssim Q_s$ during the evolution. We remove
almost all the freedom left in the solution given in Section 4, by
fixing the amplitude to be constant on the saturation line. The
$Y$-dependence of $Q_s$ is given and a multiplicative constant is
free. However, the logarithmic derivative $\lambda_s=d\log
(Q_s^2/\Lambda^2)/dY$ is well defined. In Section 6 we give our
results for $\lambda_s$ and comment on the value of $Q_s$. For
dipole-hadron scattering and at NL level in $\omega$ we find
$\lambda_s(Y=5\div 9)=0.30 \div 0.29$ and that $\lambda_s$ is
decreasing very slowly with $Y$. We also comment on the parametric
form of the boundary correction to $\lambda_s$, which is $\propto
[\bar{\alpha}_s(Q_c^2)]^{5/3}$ and well defined for any value of
$Y$, since $Q_c^2$ is increasing during the evolution. For a
suitable but reasonable choice of our free multiplicative constant
in $Q_s$, we can \underline {fix} its value to be, for example,
$Q_s(x \simeq 3 \times 10^{-4})\simeq 1 \:{\rm GeV}$. In Section 7
we give the scaling form of the amplitude on the perturbative side
of the saturation line in terms of $Q^2/Q_s^2$. This is the same
(but with a different $Q_s$) as the one obtained in leading BFKL
with either fixed or running coupling \cite{MueTri}, and therefore
we discuss its generality. In Section 8 we check our
approximations and estimate the errors which are small and under
control, and finally in Section 9 we conclude.

\section{Review of the RG improved BFKL kernel}

In this section we describe as briefly as possible the collinear
improvement of the BFKL equation. The full derivation and the
details can be found in the papers of Ciafaloni, Colferai and
Salam \cite{Salam,CiaCol,CCS,Salam2}. The BFKL equation may be
written as

\begin{equation}\label{1}
    \frac{\partial T}{\partial Y}=
    \bar{\alpha}_s\, K\otimes T,
\end{equation}

\noindent with $\bar{\alpha}_s=\alpha_s N_c/\pi$, and where we can
imagine, for example, that $T$ corresponds to the forward
scattering amplitude for a dipole of size $1/Q$ on a dipole of
size $1/\mu$. The eigenfunctions of the leading kernel are
$(Q^2/\mu^2)^{\gamma-1}$, with corresponding eigenvalues
\cite{KLF,BL}

\begin{equation}\label{2}
    \chi_0(\gamma)=
    2\psi(1)-\psi(\gamma)-\psi(1-\gamma),
\end{equation}

\noindent where $\psi(\gamma)=d\log \Gamma(\gamma)/d\gamma$. The
pole structure of $\chi_0(\gamma)$ is

\begin{equation}\label{3}
    \chi_0(\gamma)=
    \frac{1}{\gamma} + \frac{1}{1-\gamma} + {\rm finite}.
\end{equation}

\noindent The NL correction to the total characteristic function
$\chi(\gamma)$ is $\bar{\alpha}_s \chi_1(\gamma)$, where
$\chi_1(\gamma)$ is a known function \cite{FadLip,CamCia} and is
analytic in the region $0<\gamma <1$. For simplicity let us
consider its pole structure when $N_f=0$ (we will keep $N_f\neq 0$
later), which is

\begin{equation}\label{4}
    \chi_1(\gamma)=
    -\frac{1}{2 \gamma^{3}}
    -\frac{1}{2 (1- \gamma)^{3}}
    +\frac{A_{gg}(0)}{\gamma^{2}}
    +\frac{A_{gg}(0)-b}{(1-\gamma)^{2}}
    +{\rm finite}
\end{equation}

\noindent
where

\begin{equation}\label{5}
    b=\frac{\pi}{N_c} \frac{11 N_c - 2 N_f}{12 \pi},
\end{equation}

\noindent
and $A_{gg}(\omega)$ is the Mellin transform of the non-pole part of
the gluon-gluon splitting function (the pole $1/z$ has been taken into
account in the leading kernel), that is

\begin{equation}\label{6}
    A_{gg}(\omega)=\int_0^1 dz\,z^{\omega} P_{gg}(z)
    \;-\frac{1}{\omega}.
\end{equation}

\noindent When $N_f=0$ one has $A_{gg}(0)=-b=-11/12<0$, and one
can see in (\ref{4}) the large negative corrections arising from
both the double and triple poles. It was observed that the poles
at $\gamma=0$ ($\gamma=1$) arise when one considers the collinear
(anti-collinear) limit $Q^2 \gg \mu^2$ ($Q^2 \ll \mu^2$).

Since a collinear small-$x$ branching resulted in a pole $1/\gamma$ in
the leading kernel, a sequence of a small-$x$ and a non-small-$x$
branching, as can be seen from (\ref{6}) will result in a pole

\begin{equation}\label{7}
    \frac{1}{\gamma} \omega A_{gg}(\omega) \rightarrow
    \frac{1}{\gamma} \bar{\alpha}_s \chi_0(\gamma) A_{gg}(0)
    \rightarrow \frac{\bar{\alpha}_s A_{gg}(0)}{\gamma^{2}},
\end{equation}

\noindent where we have truncated the series in $\bar{\alpha}_s$
at NL level by using $\omega=\bar{\alpha}_s \chi(\gamma,\omega)$.
This is the double pole at $\gamma=0$ in Eq.(\ref{4}). Similarly
we obtain the double pole at $\gamma=1$, where there is an extra
contribution coming from the running of the coupling, if the
extracted coupling $\bar{\alpha}_s$ in Eq.(\ref{1}) is evaluated
at $Q^2$.

\noindent However, if we simply do not truncate the result in
(\ref{7}) we should obtain the all orders collinear contribution
to the BFKL kernel. Thus, including the leading kernel poles, the
total eigenvalue will become

\begin{equation}\label{8}
    \chi(\gamma,\omega)=
    \frac{1+\omega A_{gg}(\omega)}{\gamma}+
    \frac{1+\omega \left[A_{gg}(\omega)-b \right]}{1-\gamma}+
    {\rm finite},
\end{equation}

\noindent which is $\omega$-dependent. This form will be further
modified after the following discussion, since it remains to
interpret the triple poles in Eq.(\ref{4}).

Let us write the amplitude $T$ as

\begin{equation}\label{9}
    T=\int \frac{d\omega}{2\pi i}\, \frac{d\gamma}{2\pi i}\,
    T_{\omega \gamma}\, \exp \left[\omega Y - (1-\gamma) \log
    \frac{Q^2}{\mu^2}\right],
\end{equation}

\noindent where $Y=\log (s/s_0)$ and with $s$ the total energy of
the scattering. $s_0$ is an energy scale which depends on the
scales $Q$ and $\mu$ ; $s_0=s_0(Q,\mu)$. At leading level its
choice is irrelevant. At NL level one chooses $s_0$ to be the
natural scale of the problem under consideration. In the
calculation of $\omega_P$ a suitable choice is $s_0=Q \mu$. In the
collinear (anti-collinear) limit the right choice is $s_0=Q^2$
($s_0=\mu^2$). In this limit $Y=\log(1/x)=\log(s/Q^2)$, where $x$
is now the familiar Bjorken variable. As a consequence, the
eigenvalue of the NL kernel depends on $s_0$, and the one given in
Eq.(\ref{4}) is for the symmetric choice $s_0=Q\mu$. From
(\ref{9}) we can see that we can switch to $s_0=Q^2$ ($s_0=\mu^2$)
by a simple shift $\gamma \rightarrow \gamma - \frac{\omega}{2}$
($\gamma \rightarrow \gamma + \frac{\omega}{2}$), since

\begin{equation}\label{10}
    \omega \log \frac{s}{Q\mu} - (1-\gamma) \log
    \frac{Q^2}{\mu^2} \rightarrow \omega \log \frac{s}{Q^2} -
    (1-\gamma) \log \frac{Q^2}{\mu^2}.
\end{equation}

\noindent The change in the eigenvalue can be obtained from the
following iteration

\begin{align}\label{11}
    \omega & =\bar{\alpha}_s \chi(\gamma-\frac{\omega}{2})
    \rightarrow \bar{\alpha}_s \chi(\gamma) - \bar{\alpha}_s
    \frac{\omega}{2} \chi'(\gamma) \rightarrow \bar{\alpha}_s
    \chi(\gamma) - \frac{1}{2} \bar{\alpha}_s^2 \chi_0(\gamma)
    \chi'_0(\gamma)
    \nonumber \\
    & \Rightarrow \delta \chi(\gamma)= \frac{\bar{\alpha}_s}{2
    \gamma^{3}} -\frac{\bar{\alpha}_s}{2 (1-\gamma)^{3}}
    +{\rm finite},
\end{align}

\noindent where the result is again truncated at NL level. When we
add this correction to (\ref{4}) the triple pole at $\gamma=0$
cancels, as it should in the collinear limit, in order to have
consistency with the RG equations. This is the origin of the
triple poles in the NL BFKL kernel. Now one can see that
Eq.(\ref{8}) has to be substituted by

\begin{equation}\label{12}
    \chi(\gamma,\omega)=
    \frac{1+\omega A_{gg}(\omega)}{\gamma+\frac{\omega}{2}}+
    \frac{1+\omega \left[A_{gg}(\omega)-b \right]}
    {1-\gamma+\frac{\omega}{2}}+
    {\rm finite},
\end{equation}

\noindent when $s_0=Q\mu$, so that

\begin{equation}\label{13}
    \chi(\gamma,\omega)=
    \frac{1+\omega A_{gg}(\omega)}{\gamma}+
    \frac{1+\omega \left[A_{gg}(\omega)-b \right]}
    {1-\gamma+\omega}+
    {\rm finite},
\end{equation}

\noindent when $s_0=Q^2$.

It remains to combine the behavior (\ref{12}) or (\ref{13}) with
the known L and NL BFKL kernels. As explained in the next section,
the correct choice in the saturation problem is $s_0=Q^2$. In the
$\omega$-expansion approach, where one does not truncate at NL in
$\bar{\alpha}_s$ as in Eqs.(\ref{7}) and (\ref{11}), and $\omega$
appears as the expansion variable, the leading kernel is

\begin{equation}\label{14}
    \chi_0(\gamma,\omega)=
    2\psi(1)-\psi(\gamma)-\psi(1-\gamma+\omega),
\end{equation}

\noindent or

\begin{equation}\label{15}
    \chi_0(\gamma,\omega)=\chi_0(\gamma)-
    \frac{1}{1-\gamma}+\frac{1}{1-\gamma+\omega},
\end{equation}

\noindent where $\chi_0(\gamma)$ is the one given in Eq.(\ref{2}).
Both descriptions are equally good. Then we consider the NL in
$\omega$ kernel. We need to add to add in $\chi_1(\gamma)$ a term
$-\chi_0(\gamma) \chi'_0(\gamma)/2$ since $s_0=Q^2$ (see
Eq.(\ref{11})). After subtracting the NL BFKL corrections that are
already included in (\ref{14}) or (\ref{15}) to avoid double
counting, we replace all remaining poles at $\gamma=1$ with poles
at $\gamma=1+\omega$, making at the same time the substitution
$A_{gg}(0) \rightarrow A_{gg}(\omega)$ and we obtain
$\chi_1(\gamma,\omega)$. Then the NL in $\omega$ eigenvalue is

\begin{equation}\label{16}
    \chi(\gamma,\omega)=\chi_0(\gamma,\omega)+\omega\,
    \frac{\chi_1(\gamma,\omega)}{\chi_0(\gamma,\omega)},
\end{equation}

\noindent which has the proper limit given in Eq.(\ref{13}) and
includes the NL BFKL corrections. The details in constructing
$\chi_1(\gamma,\omega)$ are given in the appendix.

Our freedom in using either (\ref{14}) or (\ref{15}) for
$\chi_0(\gamma,\omega)$ (and similarly for
$\chi_1(\gamma,\omega)$) emerges from our freedom in how to shift
the poles; one can always add a finite part. However, it was shown
that any resulting difference is of order $\mathcal{O}(\omega^2)$.
This collinear resummation independence was checked in the value
of $\omega_P$ where the results differ not more than $2-3\,\%$. We
expect this to be true in the saturation problem too, since the
two kernels are very close to each other for all allowed values of
$\gamma$ and $\omega$. For practical reasons we shall use a simple
pole shift as in (\ref{15}), since in this case the
$\omega$-dependence becomes as simple as possible. Notice that
because of this $\omega$-dependence, the operator $K$ in
Eq.(\ref{1}) contains $\partial/\partial Y$ derivatives.

\section{Evolution along the critical line}

As mentioned earlier we consider the forward amplitude
$T(Q,\mu,Y)$ for dipole-dipole or dipole-hadron scattering at NL
level in $\omega$. Since the kernel eigenvalues are
$\chi(\gamma,\omega)$, the evolution equation may be written in an
operator form

\begin{equation}\label{17}
    \frac{\partial (T/\alpha_s)}{\partial Y}=
    \frac{1}{b\rho}\:\chi
    \left(1+\frac{\partial}{\partial \rho},
    \frac{\partial}{\partial Y}\right)(T/\alpha_s).
\end{equation}

\noindent Here we have defined the logarithmic variable
$\rho=\log(Q^2/\Lambda^2)$, with $\Lambda$ the usual QCD
parameter. The rapidity $Y$ is given by $Y=\log(1/x)$, with
$x=Q^2/s$ and $s$ the total energy of the scattering. It is
obvious that we have chosen an asymmetric energy scale $s_0=Q^2$.
The reason is that we are going to evolve the system, following a
particular line in the $Q^2-Y$ plane along which $Q^2$ is
increasing (we should emphasize though that the system is not in
the collinear regime). Furthermore, this is the standard
expression of Bjorken-$x$ which is also used in the HERA data. The
factor $1/b\rho$ is the running coupling $\bar{\alpha}_s$, with
$b$ from (\ref{5}).

\noindent We change variable from $\rho$ to $\eta$ according to

\begin{equation}\label{18}
    \eta=\rho-\rho_c(Y),
\end{equation}

\noindent so that

\begin{equation}\label{19}
    \frac{\partial}{\partial Y} \rightarrow
    \frac{\partial}{\partial Y}-\dot{\rho}_c
    \frac{\partial}{\partial\eta} \hspace{10pt},\hspace{10pt}
    \frac{\partial}{\partial\rho} \rightarrow
    \frac{\partial}{\partial\eta}.
\end{equation}

\noindent The dot represents a derivative with respect to $Y$.
The function $\rho_c(Y)$ will be defined soon to be a line of
almost, but not quite, constant amplitude; there will be no
exponential dependence in $Y$. $\eta$ corresponds to fluctuations
around this line. If we also expand the running coupling as

\begin{equation}\label{20}
    \frac{1}{b\rho}=
    \frac{1}{b\rho_c}-\frac{\eta}{b\rho_c^2}+
    \mathcal{O}\left(\frac{\eta^2}{\rho_c^3}\right),
\end{equation}

\noindent Eq.(\ref{17}) becomes

\begin{align}\label{21}
    b\rho_c \left(\frac{\partial}{\partial Y} -\dot{\rho}_c
    \frac{\partial}{\partial \eta} \right)&(T/\alpha_s)
    \nonumber \\
    &=\left(1-\frac{\eta}{\rho_c}\right)\;\chi
    \left(1+\frac{\partial}{\partial \eta} ,
    \frac{\partial}{\partial Y}-
    \dot{\rho}_c \frac{\partial}{\partial \eta}\right)
    (T/\alpha_s).
\end{align}

\noindent It is convenient to view the eigenvalue
$\chi(\gamma,\omega)$ as a function of $\gamma$ and $\lambda$,
where $\lambda$ is given by

\begin{equation}\label{22}
    \omega=\lambda (1-\gamma).
\end{equation}

\noindent Then, if

\begin{equation}\label{23}
    \phi(\gamma,\lambda)=
    \chi(\gamma,\lambda (1-\gamma)),
\end{equation}

\noindent Eq.(\ref{21}) may be written as

\begin{align}\label{24}
    b\rho_c \bigg(\frac{\partial}{\partial Y} - &\dot{\rho}_c
    \frac{\partial}{\partial \eta} \bigg) (T/\alpha_s)
    \nonumber \\
    &=\left(1-\frac{\eta}{\rho_c}\right)\;
    \phi \left(1+\frac{\partial}{\partial \eta},\,\dot{\rho}_c-
    \frac{\partial}{\partial Y} \left(
    \frac{\partial}{\partial \eta}
    \right)^{-1} \right)
    (T/\alpha_s).
\end{align}

\noindent We drop the $\partial/\partial Y$ in the second argument
of the $\phi$ function. This is a valid approximation that will be
checked at the end. Thus the second argument will be
$\dot{\rho}_c$. Even though $\dot{\rho}_c$ is expected to be
small, we do not expand around 0, since such an expansion would
lead us to the usual BFKL evolution, where the coefficient in the
expansion would be large at NL level. We do expand only with
respect to the first argument, in a ``diffusion approximation''
much like that introduced in \cite{CamCia2}, around the point
$\gamma_c(Y)$ to obtain

\begin{align}\label{25}
    &b\rho_c \bigg(\frac{\partial}{\partial Y} -\dot{\rho}_c
    \frac{\partial}{\partial \eta} \bigg) (T/\alpha_s)
    =\left(1-\frac{\eta}{\rho_c} \right)
    \bigg[
    \phi(\gamma_c,\dot{\rho}_c)
    \nonumber\\
    &+\phi^{1,0}(\gamma_c,\dot{\rho}_c)
    \bigg(1+\frac{\partial}{\partial \eta} - \gamma_c \bigg)
    +\frac{1}{2} \phi^{2,0}(\gamma_c,\dot{\rho}_c)
    \bigg(1+\frac{\partial}{\partial \eta} - \gamma_c \bigg)^2
    \bigg]
    (T/\alpha_s),
\end{align}

\noindent where $\phi^{m,n}$ is the $(m,n)$ derivative of
$\phi(\gamma,\lambda)$.

At this stage we need to determine the $Y$-dependence of $\gamma_c$ and
$\rho_c$. We choose them, so that $T$ is as constant as possible.
This will happen if the coefficients of the constant and
$\partial/\partial \eta$ terms in (\ref{25}) vanish. Thus

\begin{equation}\label{26}
    \phi(\gamma_c,\dot{\rho}_c)
    +(1-\gamma_c) \phi^{1,0}(\gamma_c,\dot{\rho}_c)
    -2\beta b \dot{\rho}_c=0,
\end{equation}

\noindent and

\begin{equation}\label{27}
    b \rho_c \dot{\rho}_c
    + \phi^{1,0}(\gamma_c,\dot{\rho}_c)=0.
\end{equation}

\noindent Because of the last term in (\ref{26}) the coefficient
of the constant term in (\ref{25}) does not vanish. Nevertheless,
its appearance is required in order to cancel $Y$-prefactors that
will appear in the amplitude, by a suitable choice of the pure
number $\beta$. At the end we will find $\beta=1$. Eqs.(\ref{26})
and (\ref{27}) cannot be solved in general, except in the
simplified cases of collinear models \cite{CCS2} where one keeps
only the pole parts of the characteristic function
$\chi(\gamma,\omega)$. These models could be useful in studying
the form of $Q_s(Y)$ (which is not so simple even in this case),
but its value is not very close with the one obtained when we use
the full kernel. However, one can expand around the point
$\gamma_0$, which is the solution to

\begin{equation}\label{28}
    \chi_0(\gamma_0)+(1-\gamma_0) \chi'_0(\gamma_0)=0,
\end{equation}

\noindent and with numerical value $\gamma_0=0.372...$ Notice that
(\ref{28}) is the asymptotic form of (\ref{26}) when $Y$ is very
large, since from (\ref{27}) we expect the asymptotic behavior
$\dot{\rho}_c \propto 1/\sqrt{Y}$. Therefore, a first order
expansion around the asymptotic value $\gamma_0$, simplifies our
equations to

\begin{equation}\label{29}
    \gamma_c=\gamma_0+
    \frac{-\phi(\gamma_0,\dot{\rho}_c)
    -(1-\gamma_0) \phi^{1,0}(\gamma_0,\dot{\rho}_c)
    +2 \beta b \dot{\rho}_c}
    {(1-\gamma_0) \phi^{2,0}(\gamma_0,\dot{\rho}_c)},
\end{equation}

\noindent and

\begin{equation}\label{30}
    \rho_c \dot{\rho}_c=
    \frac{\phi(\gamma_0,\dot{\rho}_c)}{b (1-\gamma_0)}
    -\frac{2 \beta \dot{\rho}_c}{1-\gamma_0}.
\end{equation}

\noindent The numerical value of
$\phi^{2,0}(\gamma_0,\dot{\rho}_c)$ is not too far from
$\chi''_0(\gamma_0)=48.5...$ The appearance of such a big number
in the denominator of (\ref{29}) suggests that the expansion
around $\gamma_0$ should be valid even in the intermediate region
$Y=6 \div 8$, which is experimentally accessible.

One can see that for very large $Y$, the scattering can be
described just by leading BFKL evolution with running coupling,
since $\dot{\rho}_c$ goes to 0. This is easily understood. During
the evolution $\rho_c$ is increasing and therefore the value of
the running coupling is decreasing. But this approach to leading
BFKL dynamics will not be realized practically, until $Y$ reaches
a very large and inaccessible value.

For what follows in the next two sections, we will not need the
explicit form of $\rho_c(Y)$ and $\gamma_c(Y)$. We return to
Eqs.(\ref{29}) and (\ref{30}) in Section 6.

\section{Solving the BFKL equation}

In this section we give an approximate solution of the RG improved
BFKL equation in the vicinity of the line $\rho_c(Y)$. Making use
of (\ref{26}) and (\ref{27}), the amplitude equation (\ref{25})
becomes

\begin{align}\label{31}
    \bigg[
    b\rho_c \frac{\partial}{\partial Y}
    -& 2\beta b \dot{\rho}_c \left(1-\frac{\eta}{\rho_c}\right)
    + \phi^{1,0}(\gamma_c,\dot{\rho}_c) \frac{\eta}{\rho_c}
    \frac{\partial}{\partial \eta}
    \nonumber\\
    -& \frac{1}{2} \phi^{2,0}(\gamma_c,\dot{\rho}_c)
    \left(1-\frac{\eta}{\rho_c}\right)
    \left(1+\frac{\partial}{\partial \eta} - \gamma_c \right)^2
    \bigg]
    (T/\alpha_s)=0.
\end{align}

\noindent When $\eta \ll \rho_c$ one can drop the corresponding
term in the first parenthesis of the last term. Now let's isolate
the leading exponential, in the scaling variable $\eta$, behavior
of the amplitude and write $T$ as

\begin{equation}\label{32}
    T=\alpha_s \exp[-(1-\gamma_c) \eta +2 \beta \log\rho_c]
    \psi(\eta,Y).
\end{equation}

\noindent We have neglected any target dependent ($\mu$-dependent)
prefactors in $T$, which are proportional to $1/\mu^2$.
We easily find that $\psi$ satisfies

\begin{align}\label{33}
    \bigg \{
    -&\phi^{1,0}(\gamma_c,\dot{\rho}_c)
    \frac{\partial}{\partial \rho_c}
    + \phi^{1,0}(\gamma_c,\dot{\rho}_c) \frac{\eta}{\rho_c}
    \frac {\partial}{\partial \eta}
    \nonumber\\
    -&\frac{1}{2} \phi^{2,0}(\gamma_c,\dot{\rho}_c)
    \frac{\partial^2}{\partial \eta^2}
    +[\phi(\gamma_c,\dot{\rho}_c)+b\rho_c^2\dot{\gamma}_c]
    \frac{\eta}{\rho_c}
    \bigg \} \psi=0,
\end{align}

\noindent which becomes now our basic equation. Since there is no
explicit $Y$ dependence in (\ref{33}), we have decided to view the
function $\psi$ as a function of $\rho_c$ instead of $Y$,
$\psi=\psi(\eta,\rho_c)$, by means of the transformation

\begin{equation}\label{34}
    \frac{\partial}{\partial Y} \rightarrow \dot{\rho}_c
    \frac{\partial}{\partial \rho_c},
\end{equation}

\noindent and we have simplified as much as possible the
coefficients by using Eqs.(\ref{26}) and (\ref{27}).

Formally these coefficients should be expressed in terms of
$\rho_c$, which can be done, in principle, when one uses
Eqs.(\ref{29}) and (\ref{30}). Practically it is easier to have
them as functions of $\lambda_c \equiv \dot{\rho}_c$. In any case,
even though these are complicated functions, we will need only
their asymptotic values in the approximate solution of (\ref{33}).

It is convenient to make a final change of variables from
$(\eta,\rho_c)$ to $(\xi,t)$, where

\begin{equation}\label{35}
    \xi=\frac{\eta}{D \rho_c^{1/3}},
\end{equation}

\noindent and

\begin{equation}\label{36}
    t=3 D_0 (1-\gamma_0) \rho_c^{1/3},
\end{equation}

\noindent with

\begin{equation}\label{37}
    D=\left\{\frac{\phi^{2,0}(\gamma_c,\dot{\rho}_c)}
    {2[\phi(\gamma_c,\dot{\rho}_c)+b\rho_c^2\dot{\gamma}_c]}
    \right\}^{1/3},
\end{equation}

\noindent and

\begin{equation}\label{38}
    D_0=\left[ \frac{\chi''_0(\gamma_0)} {2\chi_0(\gamma_0)}
    \right]^{1/3}.
\end{equation}

\noindent Clearly $D_0=1.99...$ is the asymptotic value of $D$.
After some straightforward algebra Eq.(\ref{33}) becomes

\begin{equation}\label{39}
    \left[
    h_1(t) \frac{\partial}{\partial t}
    -4 h_2(t) \frac{\xi}{t} \frac{\partial}{\partial \xi}
    -\frac{\partial^2}{\partial \xi^2}
    +\xi
    \right] \psi=0.
\end{equation}

\noindent While the coefficients of the last two terms have been
totally simplified, the ones $h_1(t)$ and $h_2(t)$ of the first
two terms are extremely complicated functions, given by

\begin{equation}\label{40}
    h_1(t)=-\frac{D_0 \,(1-\gamma_0)
    \phi^{1,0}(\gamma_c,\lambda_c)}
    {D \, \phi(\gamma_c,\lambda_c)},
\end{equation}

\noindent and

\begin{equation}\label{41}
    h_2(t)=h_1(t)
    \left[
    1+\frac{3 D'(\lambda_c) \rho_c(\lambda_c)}
    {4 D \rho'_c(\lambda_c)}
    \right].
\end{equation}

\noindent As said before, $h_1$ and $h_2$ have been expressed in
terms of $\lambda_c=\dot{\rho}_c$. The prime stands for a
derivative with respect to $\lambda_c$, and $\gamma_c$ and
$\rho_c$ are given as functions of $\lambda_c$ from Eqs.(\ref{29})
and (\ref{30}). $\dot{\gamma}_c$ which is contained in the
definition of $D$, can also be easily expressed as a function of
$\lambda_c$ through a chain rule differentiation giving
$\dot{\gamma}_c=\gamma'_c(\lambda_c) \lambda_c
/\rho'_c(\lambda_c)$. In the asymptotic region of large $t$, or
equivalently of small $\lambda_c$, $h_1$ and $h_2$ approach 1,

\begin{equation}\label{42}
    \lim_{t\to\infty}h_1(t)
    =\lim_{t\to\infty}h_2(t)=1,
\end{equation}

\noindent with corrections of order $\mathcal{O}(1/t^3)$, which
turn out to be small not only parametrically, but even numerically
for intermediate values of $Y$.

It appears difficult to find an exact solution of Eq.(\ref{39})
even if we set $h_1=h_2=1$. However, we can find an approximate
solution of the form

\begin{equation}\label{43}
    \psi(\xi,t)={\rm Ai}(\xi-\sigma)
    \exp
    \left[
    -\frac{\xi^2}{t} h_2(t)
    -\sigma \int \frac{dt}{h_1(t)}
    -2 \int \frac {dt\,h_2(t)}{t\,h_1(t)}
    \right],
\end{equation}

\noindent where $\sigma$ is a constant to be determined and Ai is
the Airy function. This expression satisfies Eq.(\ref{39}) up to a
term $R(\xi,t) \psi$, where

\begin{equation}\label{44}
    R(\xi,t)=\frac{\xi^2}{t^2}
    \left[
    4 h_2^2 + h_1 h_2 - t h_1 \frac{dh_2}{dt}
    \right]
    =\frac{5 \xi^2}{t^2}
    \left[
    1 + \mathcal{O} \left( \frac{1}{t^3} \right)
    \right].
\end{equation}

\noindent It is important that this remainder contains no terms of
the order $f(\xi)/t$, as those terms would be important when one
integrates (\ref{39}) over $t$. The approximate solution given,
should be the one that matches the exact solution in the
intermediate and asymptotic region. It contains the Airy function
\cite{Lipatov} and the diffusion type factor $\exp(-\xi^2/t)$,
which appear in general when one considers BFKL evolution with
running coupling. We also mention that at this point we loose
track of the overall constant which appears in front of the
amplitude, as a consequence of our inability to solve (\ref{39})
exactly.

\section{The saturation boundary}

Up to target dependent factors, the amplitude $T$ is given by
Eq.(\ref{32}) with $\psi$ as found in Eq.(\ref{43}). It contains
two parameters $\beta$ and $\sigma$, which are uniquely determined
from the saturation boundary conditions in the asymptotic region.
To that end we express $T$ in terms of the variables $(\xi,t)$
in the asymptotic region, that is

\begin{equation}\label{45}
    T={\rm Ai}(\xi-\sigma)
    \exp \left[ -\frac{\xi t}{3} - \sigma t
    + (6 \beta -5) \log t \right].
\end{equation}

\noindent When $\xi$ becomes negative, the amplitude grows
strongly, it reaches a maximum value and then it decreases rapidly
to vanish at $\xi=\xi_1 + \sigma$, where $\xi_1=-2.33...$ is the
first zero of the Airy function. A decreasing amplitude is of
course an unphysical situation, which simply means that we loose
our ability to describe the evolution of the system by using
linear equations. Thus we need to put a saturation boundary before
$T$ hits its maximum value, along which we fix $T$ to be constant,
in order to illustrate the effects of the non-linear terms in the
simplest possible way. Now writing $\xi$ as

\begin{equation}\label{46}
    \xi=\xi_1+\sigma+\delta\xi,
\end{equation}

\noindent we have

\begin{equation}\label{47}
    {\rm Ai}(\xi-\sigma)=
    {\rm Ai'}(\xi_1)
    \left[
    \delta \xi
    +\frac{\xi_1}{6} \delta \xi^3
    +\frac{1}{12} \delta \xi^4
    +\mathcal{O}(\delta \xi^5)
    \right],
\end{equation}

\noindent and we keep only the linear term as we will shortly see
that $\delta \xi$ is small in the region of interest. Then
Eq.(\ref{45}) becomes (dropping again unimportant for our purposes
multiplicative constants)

\begin{equation}\label{48}
    T=\exp \left[
    -\frac{(\xi_1+4 \sigma)t}{3}+
    (6 \beta -5) \log t
    -\frac{\delta \xi \, t}{3} +\log \delta \xi
    \right].
\end{equation}

\noindent The maximum of this expression occurs
when $\delta \xi=3/t$, thus
we assume a saturation boundary at

\begin{equation}\label{49}
    \delta \xi_s= \frac{3 (1+c)}{t},
\end{equation}

\noindent where the amplitude is a finite fraction of its maximum
value. $c$ is a positive constant of order $\mathcal{O}(1)$, that
we are not able to determine. On the boundary the amplitude is

\begin{equation}\label{50}
    T_s= \exp \left[
    -\frac{(\xi_1+4 \sigma)t}{3} +6( \beta -1)\log t
    \right].
\end{equation}

\noindent Clearly, we need to choose our two parameters to be

\begin{equation}\label{51}
    \beta=1 \hspace{10pt}, \hspace{10pt} \sigma=-\frac{\xi_1}{4},
\end{equation}

\noindent so that $T$ is well defined
and $t$-independent on the boundary.

Now we are finally in a position to give an expression for the
saturation momentum. Using the results stated in Eqs.(\ref{49}) and
(\ref{51}) and recalling the definitions (\ref{18}), (\ref{35}),
(\ref{36}) and (\ref{46}) we obtain

\begin{equation}\label{52}
    \rho_s=\rho_c+\frac{3 \xi_1}{4}\, D_0 \, \rho_c^{1/3}+
    \frac{1+c}{1-\gamma_0},
\end{equation}

\noindent which gives

\begin{equation}\label{53}
    Q_s=\Lambda \exp \left[
    \frac{\rho_c}{2} +
    \frac{3 \xi_1}{8}\, D_0 \, \rho_c^{1/3} +
    \frac{1+c}{2 (1-\gamma_0)}
    \right].
\end{equation}

\noindent where we have replaced $D$ by its asymptotic value
$D_0$. It is obvious that in our approach there is a free
multiplicative constant. However, for the logarithmic derivative
of the saturation momentum, $\lambda_s=\dot{\rho}_s$, there is no
such ambiguity. In Section 8 we will discuss possible
uncertainties in Eqs.(\ref{52}) and (\ref{53}), as far as the
energy dependence is considered, which turn out to be
parametrically and numerically small. In order to make use of
(\ref{52}) or (\ref{53}) we have to determine $\rho_c(Y)$, a task
that we now turn to.

\section{Results for $\lambda_s$}

Given a kernel, we need to solve Eq.(\ref{30}) to find the
$Y$-dependence of $\rho_c$. In general it appears easier to solve
(when a solution is possible) for $\lambda_c=\dot{\rho}_c$.
Considering  $\rho_c$ as a function of $\lambda_c$ and 
differentiating with respect to $Y$ using the chain rule, 
we have

\begin{equation}\label{55}
    \lambda_c=\rho'_c(\lambda_c) \frac{d \lambda_c}{dY},
\end{equation}

\noindent and by separating variables we obtain 

\begin{equation}\label{56}
    \int_{\lambda_c(0)}^{\lambda_c(Y)} d\lambda_c\,
    \frac{\rho'_c(\lambda_c)}{\lambda_c}=Y,
\end{equation}

\noindent where 

\begin{equation}\label{54}
    \rho_c(\lambda_c)=
    \frac{\phi(\gamma_0,\lambda_c)}{b (1-\gamma_0) \lambda_c}
    -\frac{2}{1-\gamma_0},
\end{equation}

\noindent as determined by Eq.(\ref{30}).

\noindent It is clear that one needs an initial condition
specified in order to proceed. In leading BFKL with fixed
coupling, the initial condition is totally irrelevant for
$\lambda_c$ since the leading behavior of $\rho_c(Y)$ in that case
is purely linear. In the running coupling case, either in BFKL or
in $\omega$-expansion, there is a an integration constant $Y_0$ or
$\lambda_c(0)$ as it appears in (\ref{56}), which affects the
value of $\lambda_c(Y)$. However, as we discuss shortly after and
in Section 8, small changes in the initial conditions will
not change our results significantly and furthermore this change
is predictable. The most natural initial condition to adopt is

\begin{equation}\label{57}
    \rho_c(Y=0)=\log \frac{\mu^2}{\Lambda^2},
\end{equation}

\noindent with $\mu^{-1}$ of the order of the target size. Then it
is straightforward to find $\lambda_c(0)$ from (\ref{54}). Now
after determining $\lambda_c$ and/or $\rho_c$, one can finally get
the logarithmic derivative of the saturation momentum from the
differentiation of (\ref{52}), which gives

\begin{equation}\label{58}
    \lambda_s=\lambda_c+\frac{\xi_1 D_0}{4 \rho_c^{2/3}} \lambda_c.
\end{equation}

\noindent This is our main result for $\lambda_s$. In the following
we examine the cases of Leading BFKL, Leading in $\omega$ and NL
in $\omega$ dynamics.

\subsection{Leading BFKL - review}

In this case it is straightforward to solve directly for
$\rho_c(Y)$ using Eq.(\ref{30}). We find

\begin{equation}\label{59}
    \rho_c(Y)=\sqrt{
    \frac{2 \chi_0(\gamma_0)}{b (1-\gamma_0)}Y
    +\left[\rho_c(0)+\frac{2}{1-\gamma_0} \right]^2}
    -\frac{2}{1-\gamma_0},
\end{equation}

\noindent from which it is trivial to obtain $\lambda_c(Y)$. Then
(\ref{58}) provides us with an analytical expression for
$\lambda_s(Y)$.

\noindent It is very useful to examine two particular limits of
Eq.(\ref{59}), in order to get a better understanding of the
initial conditions. \vspace{6pt}

\noindent {\bf (i)} Let's imagine first that the target is a very
small dipole, such that the second term of the square root in
(\ref{59}) dominates, that is

\begin{equation}\label{60}
    \rho_c^2(0) \gg
    \frac{2 \chi_0(\gamma_0)}{b (1-\gamma_0)}Y.
\end{equation}

\noindent Then one can easily see that (\ref{59}) reduces to the
exact fixed coupling result

\begin{equation}\label{61}
    \rho_c(Y)=\rho_c(0) +
    \frac{\chi_0(\gamma_0)}{b (1-\gamma_0) \rho_c(0)}Y,
\end{equation}

\noindent with $\rho_c(0)$ as given in (\ref{57}) and the natural
identification $1/b \rho_c(0) \rightarrow \bar{\alpha}_s$. This is
expected since (\ref{60}) simply states the fact that, the
rapidity is not large enough (as $1/\alpha_s^2$ parametrically) and
therefore the NL and the running coupling corrections are
unimportant. \vspace{6pt}

\noindent{\bf (ii)} Now let's consider that $Y$ is large enough so
that the first term of the square root dominates, that is

\begin{equation}\label{62}
    \rho_c^2(0) \ll
    \frac{2 \chi_0(\gamma_0)}{b (1-\gamma_0)}Y.
\end{equation}

\noindent Then (\ref{59}) reduces to

\begin{equation}\label{63}
    \rho_c(Y)=\sqrt{
    \frac{2 \chi_0(\gamma_0)}{b (1-\gamma_0)}Y},
\end{equation}

\noindent In this case, where the running coupling dynamics is
necessary, it is clear that the initial conditions are lost in the
region that (\ref{62}) is satisfied. Parametrically, the
uncertainty in $\lambda_c(Y)$, due to the initial conditions,
vanishes as $1/Y^{3/2}$.\vspace{5pt}

Thus it becomes crucial to specify the initial
condition, e.g. whether the target is a small dipole or a hadron,
and the region of $Y$ values that one is interested in, before
making any kind of expansion in (\ref{59}). Even though we would
like to imagine in general that $Y$ is large, let's consider a
realistic case where the target is of size $\mu \approx\, 1 {\rm
GeV}$ and the rapidity is in the region $Y=6\div 8$. It is easy to
check that the two terms in (\ref{59}) are of comparable size.
Therefore we use this equation without making any approximations.
But there are two important points emerging from the above
analysis, which will also be true in the case of the $\omega$
expansion. The first point is that a running coupling description
has to be considered in a correct (at least conceptually)
treatment of the saturation problem. The second point is that the
initial condition is not lost in our case, that is the result
depends on $\rho_c(0)$, but any small changes in the initial
conditions are almost unobservable since they parametrically vanish
as $1/\rho_c^{3/2}$ when one considers $\lambda_c$.

\subsection{Leading in $\omega$}

We consider the kernel given in Eq.(\ref{15}). The $\phi$ function
is

\begin{equation}\label{64}
    \phi(\gamma,\lambda)=
    \chi_0(\gamma)
    -\frac{\lambda}{(1-\gamma)(1+\lambda)}.
\end{equation}

\noindent The $\rho_c$ equation, Eq.(\ref{30}), becomes

\begin{equation}\label{65}
    \rho_c \dot{\rho}_c=
    \frac{\chi_0(\gamma_0)}{b (1-\gamma_0)}
    -\frac{\dot{\rho}_c}{b (1-\gamma_0)^2 (1+\dot{\rho}_c)}
    -\frac{2 \dot{\rho}_c}{1-\gamma_0}.
\end{equation}

\noindent Even though $\dot{\rho}_c$ is expected to be small, we
prefer not to do any kind of expansion in the second term of the
right-hand side, in order to be consistent with the
$\omega$-expansion approach. If we expand this term, we loose at
the same time the correct collinear behavior as explained in
Section 2. We also notice the negative sign of this term, which is
expected, since the leading in $\omega$ kernel contains a part of
the NL BFKL kernel. Now, following the general strategy presented
in the beginning of this section, we can do the integral in
(\ref{56}) to obtain

\begin{equation}\label{66}
    \frac{\chi_0(\gamma_0)}{2 b (1-\gamma_0) \lambda_c^2}
    -\frac{1}{b (1-\gamma_0)^2}
    \left(
    \log \frac{1+\lambda_c}{\lambda_c}
    -\frac{1}{1+\lambda_c}
    \right)
    -[\lambda_c \rightarrow \lambda_c(0)]=Y.
\end{equation}

\noindent Even though we cannot analytically solve this equation for
$\lambda_c$ (we would like to avoid an iterative procedure for
reasons that we explained before), this is a transcendental
equation that can be solved numerically for a given $Y$. Then we
proceed to determine $\rho_c$ from (\ref{54}) and $\lambda_s$ from
(\ref{58}).

We mention here that if we use the ``equivalent'' kernel given in
(\ref{14}), we will not be able to reach a result as simple as the
one stated in (\ref{66}), as it appears difficult to integrate
Eq.(\ref{56}). However the two kernels should give the same
results.

\subsection{Next to leading in $\omega$ }

The kernel in this case is the one in (\ref{16}), where
$\chi_1(\gamma,\omega)$ is a known function that is constructed in
the Appendix following the procedure of
Refs.\cite{Salam,CiaCol,CCS,Salam2}. Here it is not possible to
give a relatively simple expression as Eq.(\ref{66}) in the
previous subsection. The reason is that $\chi_1(\gamma,\omega)$
contains a piece $A_{gg}(\omega)$, the Mellin transform of the
gluon splitting function. This has a term $\psi(2+\omega)$ and
Eq.(\ref{56}) cannot be integrated . In the region of interest the
expected value of $\omega$ is $\omega=\lambda_c (1-\gamma_c)$,
which can be found (a posteriori) to be around $0.25$. Thus we can
expand $A_{gg}(\omega)$, or just its $\psi(2+\omega)$ part, around
$\omega=0$. Then, depending on the order of this Taylor expansion,
we find a more and more complicated transcendental equation (even
at 0th order) containing fractions of polynomials and logarithms
of $\lambda_c$. This equation is the analogous of (\ref{66}), that
we found in the leading in $\omega$ case. There is no reason to
present such a complicated expression here, but it is again an
algebraic equation which can be solved numerically. We find, for a
given $Y$, that $\lambda_c$ reaches rapidly a fixed value as we
increase the order of the expansion mentioned above; even if we
keep just the 0th order term, the result is very close to the
exact value, something which is expected since we have
approximated only a small piece of the NL in $\omega$ kernel. In
our results for $\lambda_s$, that appear in the table of the next
subsection, we have kept enough terms in this expansion so that
the obtained value of $\lambda_c$ is practically the exact one.
$\lambda_s$ is obtained in the way we have already explained in
the previous subsections.

There is another, but clearly approximate, iterative procedure,
that we can follow in order to obtain a simple equation, which
nevertheless gives the same results for the NL in $\omega$
corrections. We write the full kernel as

\begin{equation}\label{67}
    \chi(\gamma,\omega)=
    \chi_0(\gamma,\omega)
    +\omega \,\delta,
\end{equation}

\noindent and treat the term $\omega \, \delta$ as a perturbation
in the following sense. For a given $Y$ we solve the leading in
$\omega$ problem. Then for the resulting value of $\lambda_c$ we
determine the number $\delta$ appearing in (\ref{67}) from
$\delta=\chi_1(\gamma_c,\omega_c)/\chi_0(\gamma_c,\omega_c)$,
where $\omega_c=\lambda_c (1-\gamma_c)$ as dictated from
(\ref{22}). Then we insert this value of $\delta$ in (\ref{67})
and solve the new problem, which is of equal difficulty as the
leading in $\omega$ problem. We obtain a new value of $\lambda_c$
and we can repeat the procedure. This converges rapidly and after
a couple of iterations $\delta$ and $\lambda_c$ become constant.
Then we can repeat for different $Y$. We have decided to
extract explicitly a factor of $\omega$ in (\ref{67}), instead of
treating all the NL correction as a constant, since this factor
characterizes the very leading behavior of the NL term.

\subsection{Summary of results and discussion}

We assume an initial condition $\rho_c(0)=2 \log 5$, so that the
target is of size $\mu^{-1} \approx 1 \:{\rm GeV}^{-1}$, when
$\Lambda \approx 200 \:{\rm MeV}$. We set the number of active
flavors to be $N_f=3$. In the table we summarize the values of
$\lambda_s$, obtained for the different kernels considered and for
various values of $Y$. \vspace{15pt}

\begin{center}
\begin{tabular}{|r||c|c|c||c|} \hline
 \multicolumn{5}{|c|}
 {\textsf{Values of $\lambda_s=d\log(Q_s^2/\Lambda^2)/dY$}}
 \\ \hline
  $Y$ & BFKL & L in $\omega$ & NL in $\omega$ &
  $\delta \chi /\chi_0$ \\
  \hline\hline
  5 & 0.433 & 0.366 & 0.300 & -0.20 \\ \hline
  6 & 0.419 & 0.358 & 0.297 & -0.20 \\ \hline
  7 & 0.406 & 0.350 & 0.294 & -0.19 \\ \hline
  8 & 0.393 & 0.343 & 0.291 & -0.19 \\ \hline
  9 & 0.382 & 0.336 & 0.288 & -0.18 \\ \hline
 ... & ... & ... & ... & ...     \\ \hline
  25 & 0.278 & 0.258 & 0.238 & -0.13 \\
  \hline
\end{tabular}
\end{center}
\vspace{15pt}

\noindent Some discussion and a couple of comments need to follow
here. \vspace{6pt}

\noindent {\bf (i)} In all cases, and particularly in the $\omega$
expansion approach, $\lambda_s$ is decreasing very slowly. We have
included the inaccessible value $Y=25$ in the table, just to
illustrate this slow decrease. In the case of the NL in $\omega$
kernel, its value is practically constant in a certain $Y$ region.
For example, in the region of phenomenological interest $10^{-2}
\gtrsim x \gtrsim 10^{-4}$, which corresponds to $5 \lesssim Y
\lesssim 9$, we effectively have $\lambda_s=0.30 \div 0.29$ and
therefore the energy dependence of the saturation momentum is well
described by the simple exponential law $Q_s^2(Y) \propto
\Lambda^2 \exp (\lambda_s Y)$. This is in agreement with
saturation based models \cite{GBW} that have successfully
interpreted much of the HERA data. In these models such an ansatz
for $Q_s$ was taken, with $\lambda_s$ as a free parameter, and
after the fitting to the data was performed, the resulting value
of $\lambda_s$ was very close to the one we quoted above.

\noindent However, we should emphasize again that a fixed coupling
description (which gives a constant value for $\lambda_s$, but
much larger than the one found here) is not the right approach to
the saturation problem, at least conceptually. The variation of
$\lambda_s$ is rather slow partly because of the particular
interplay between the leading behavior and the saturation
corrections, as one can check in Eq.(\ref{58}), and partly because
of NL corrections which slow down the evolution.

\noindent Now we come to the issue of the actual value of $Q_s$.
As already mentioned at the end of Section 5, we do have control
of the energy dependence of $Q_s$, but there is an ambiguity of a
free multiplicative constant as can be seen in Eq.(\ref{53}). Thus
we cannot give any accurate prediction for the value of $Q_s$. We
just mention though, that for a ``reasonable'' choice of $c$ in
(\ref{53}), we obtain a phenomenologically accepted result. For
example, with $c=3/4$ (and $\Lambda=200 \:{\rm MeV}$), we find
$Q_s(x \simeq 3 \times 10^{-4})\simeq 1 \:{\rm GeV}$.

\noindent In a very recent work \cite{GLLM} a fit to the HERA data
has been performed. The model relies on the combination of the
Balitsky-Kovchegov and the DGLAP equations. In such a case NL
effects are neglected. As a part of their results, a best fit
value of $\lambda_s=0.36 \pm 0.04$ is given, where the form $Q_s^2
\propto \Lambda^2 \exp (\lambda_s Y)$ has been assumed. It is
interesting that the central value of this result coincides with
our Leading in $\omega$ one.\vspace{6pt}

\noindent {\bf (ii)} It is clear that as $Y$ increases the results
of the $\omega$ expansion converge to those of leading BFKL. This
is expected as already mentioned at the end of Section 3. But this
kind of convergence is slow, and it will not happen until $Y$
reaches a very large value. Thus, in the region of
phenomenological interest, the effects of the collinear
improvement of the BFKL kernel are important. The leading BFKL
prediction is reduced by an amount around $25-30 \%$ when NL in
$\omega$ corrections are included.

\noindent It is also instructive to compare the leading and the NL
in $\omega$ results. In the last column of the table we give the
ratio $\delta \chi /\chi_0$, which is the NL correction to the
kernel divided by the leading one and evaluated at the point
$(\gamma_c,\omega_c)$. We notice that the correction has a
``natural size'' of the order of $\mathcal{O}(\alpha_s)$, when
compared to the leading contribution, where the running $\alpha_s$
is evaluated at the relevant value of $Q^2$, say somewhere between
$Q_c^2(Y)$ and $Q_s^2(Y)$. Essentially this is a key element of
the RG improvement of the BFKL equation. NL corrections are
naturally small. We mention here that this observation is also
true in the problem of the calculation of the hard Pomeron intercept
$\omega_P$ \cite{CiaCol,CCS}.\vspace{6pt}

\noindent {\bf (iii)} In all cases, the second term in (\ref{58})
which represents the effects of the saturation boundary, is
smaller in magnitude than the leading term, but its contribution
is extremely important. Even though $\rho_c$ is big, this term
will not be negligible until $Y$, and therefore $\rho_c$, reaches
a huge value.

It is quite interesting to investigate the parametric form of this
saturation correction. In view of Eq.(\ref{22}) we naturally
define $\omega_c=(1-\gamma_0)\lambda_c$ and
$\omega_s=(1-\gamma_0)\lambda_s$. Expressing $\lambda_c$ in terms
of $\rho_c$ in the 2nd term of (\ref{58}), by keeping only the
very leading behavior as determined from the dominant term in
(\ref{65}), we obtain

\begin{align}\label{100}
    \omega_s&=\omega_c
    +\frac{\xi_1 D_0 \chi_0(\gamma_0)}
    {4 b} \frac{1}{\rho_c^{5/3}}
    \nonumber \\
    &=\omega_c+\frac{\xi_1}{4}
    \bigg[\frac{b^2 \chi_0^2(\gamma_0)
    \chi''_0(\gamma_0)}{2}\bigg]^{1/3}
    [\bar{\alpha}_s(\rho_c)]^{5/3}.
\end{align}

\noindent Contributions of the type $\bar{\alpha}_s^{5/3}$ due to
the running of the coupling have been observed and studied in the
calculation of the hard Pomeron intercept
\cite{KovMue,ABB,Levin,CMT,CCS}. There is a striking similarity
between the above equation and Eq.(4.9) in \cite{CCS}. In our case 
the $\chi$ function is evaluated at $\gamma_0$ and not
at its minimum, and the additional $1/4$ factor is understandable
through Eq.(\ref{51}).

\noindent We need to mention though, that in the saturation
problem there is no limit in the applicability of (\ref{58}) or
(\ref{100}). On the contrary, since the line of evolution is one
of increasing $\rho_c$ and decreasing $\bar{\alpha}_s$, the second
term is always smaller than the leading one.

\section{The amplitude and geometrical scaling}

In this section we return to discuss the form of the amplitude.
Since this is constant along the saturation line, we naturally
expect a scaling behavior in a certain region around this line. 
Of course we restrict ourselves to the perturbative side of the
boundary, since we do not know how to handle the details of the
non-linear effects. The scaling form of the amplitude should be
more and more exact as $Y$ increases, but it is also a reliable
expression even in the region of phenomenological interest.

Returning to Eq.(\ref{45}), with $\beta$ and $\sigma$ as obtained
in (\ref{51}), and expressing the variables $(\xi,t)$ in terms of
the more physical variables $(Q,Q_s)$, following the definitions
we have given throughout this paper, it is straightforward to find

\begin{equation}\label{68}
    T \propto
    \left(\frac{Q_s^2}{Q^2}\right)^{1-\gamma_0}
    \left[ \log \frac{Q^2}{Q_s^2}
    + \frac{1+c}{1-\gamma_0}\right].
\end{equation}

\noindent The target dependent factors that are not written in
this expression are of the form (${\rm const.}/\mu^2$), with $\mu$
characterizing as always the target size. We notice also that the
form of the amplitude will not change if we change $Q_s$ by a
multiplicative constant, since such an operation can be absorbed
in a redefinition of the constant $c$. Eq.(\ref{68}) is valid in
the region

\begin{equation}\label{69}
    \log \frac{Q^2}{Q_s^2} \lesssim
    \sqrt{\frac{6}{|\xi_1|}} D_0 \rho_c^{1/3}
    -\frac{1+c}{1-\gamma_0},
\end{equation}

\noindent a constraint arising from the expansion of the Airy
function. Parametrically this constraint is satisfied since
$\rho_c$ is growing with $Y$. In the region of interest, the
number on the right hand side is not large enough to allow scaling
in a wide region of $Q^2$ values, as it has been observed in the
HERA data \cite{GBKS}. Even at the very small value $x \simeq
10^{-4}$, the inequality in (\ref{69}) will totally break at $Q^2
\gtrsim 50 \:{\rm GeV}^2$ (independent of the unknown constant
$c$). This ``scaling'' window depends relatively strongly on the
initial condition. Furthermore, a more complete solution of
(\ref{39}) could also change the above estimate, but we do not
know how to control this issue in a more precise way. Other
approximations that we have made, for example $\eta \ll \rho_c$,
are also parametrically satisfied and are much weaker numerically.

Eq.(\ref{68}) is the same (but with a different $Q_s$) as the one
derived in Ref.\cite{MueTri}, in a leading BFKL description of the
saturation problem with either fixed or running coupling.
Therefore it is interesting to try and interpret this result. The
power behavior, with the anomalous dimension $1-\gamma_0$, is a
consequence of the BFKL dynamics, either the ``standard'' or the
collinearly improved. The logarithm in the square bracket is the
effect of the absorptive boundary. Such a boundary imposes the
condition that somewhere beyond the saturation line (where of
course (\ref{68}) is not valid any more since non-linear effects
are important) the amplitude vanish. Clearly a logarithm satisfies
this condition. We can give a more mathematical argument that
(\ref{68}) is the only scaling form allowed by the dynamics we
have considered.

To illustrate the issue in a simple way let us consider fixed
coupling leading BFKL dynamics. In this case Eq.(\ref{17})
simplifies to

\begin{equation}\label{70}
    \frac {\partial T}{\partial Y}
    =\bar{\alpha}_s \,\chi_0
    \left(1+\frac{\partial}{\partial \rho}\right)T.
\end{equation}

\noindent Following a similar procedure as in Section 3, we define
now $\eta=\rho-\rho_s(Y)$ and  write the amplitude as
$T=\exp[-(1-\gamma_0) \eta]\,\psi(\eta)$ to obtain

\begin{equation}\label{71}
    \left[
    \bar{\alpha}_s \, \chi_0
    \left(\gamma_0+\frac{\partial}{\partial \eta} \right)
    -\dot{\rho}_s (1-\gamma_0)
    +\dot{\rho}_s \frac{\partial}{\partial \eta}
    \right] \psi(\eta)=0,
\end{equation}

\noindent with $\gamma_0$ a constant undetermined for the moment
and where we dropped the $\partial/\partial Y$ term in order to
look for an exact scaling solution. Clearly $\dot{\rho}_s$ has to
be constant. Expanding the $\chi_0$ function around $\gamma_0$,
(\ref{71}) becomes

\begin{align}\label{72}
    \bigg\{
    \bar{\alpha}_s \chi_0(\gamma_0) - (1-\gamma_0) \dot{\rho}_s
    & +[\bar{\alpha}_s \chi'_0(\gamma_0)+\dot{\rho}_s]
    \frac{\partial}{\partial \eta}
    \nonumber\\
    & +\bar{\alpha}_s \sum_{n=2}^{\infty}
    \frac{\chi_0^{(n)}(\gamma_0)}{n!}
    \left(\frac{\partial}{\partial \eta}\right)^n
    \bigg\}\psi(\eta)=0.
\end{align}

\noindent Let us assume that that $\psi(\eta)$ is a finite
polynomial of $\eta$ and of order $k$, that is

\begin{equation}\label{73}
    \psi(\eta)=a_0+ a_1 \eta +a_2 \eta^2 +...+a_k \eta^k.
\end{equation}

\noindent It is easy to see that the coefficients of the $\eta^k$
and $\eta^{k-1}$ terms in (\ref{72}) will vanish if and only if

\begin{equation}\label{74}
    \bar{\alpha}_s \chi_0(\gamma_0) - (1-\gamma_0) \dot{\rho}_s=0,
\end{equation}

\begin{equation}\label{75}
    \bar{\alpha}_s \chi'_0(\gamma_0)+\dot{\rho}_s=0,
\end{equation}

\noindent respectively and $\gamma_0$ becomes our usual one. But
if $k\not=0,1$ the coefficients of the remaining terms will not
vanish and (\ref{72}) is not satisfied. Thus only the case $k=1$,
which is what we have found in (\ref{68}), is allowed. The case
$k=0$ is allowed too, but it is not consistent with our saturation
boundary conditions. Notice that (\ref{74}) and (\ref{75}) are
equivalent to the reduction of (\ref{26}) and (\ref{27}) in the
fixed coupling BFKL case, with $\gamma_c \rightarrow \gamma_0$ and
$\beta \rightarrow 0$.

The above can be generalized in the case of an RG improved kernel
with running coupling in a straightforward way. Then the solution
with $k=1$ is an approximate one when $\eta \ll \rho_s$ and we are
lead to Eqs.(\ref{26}) and (\ref{27}), again with $\gamma_c
\rightarrow \gamma_0$ and $\beta \rightarrow 0$. The one is
satisfied in the asymptotic region and the other determines the
leading behavior of $\rho_s$. Notice that this procedure gives us
a very fast way to obtain (\ref{68}), but only the leading
behavior of $Q_s$, which is $Q_c$, is recovered. That is, we do
not find in $Q_s$ the corrections due to the boundary, which
turned out to be important as we saw in the previous sections.

\section{Estimation of errors}

Here we check the approximations that we have made and which could
be possible sources of errors. We find that all these errors are
small and under control. \vspace{6pt}

\noindent {\bf (i)} There is an ambiguity arising from the initial
condition. $\rho_c(0)$ should characterize the target size and is
given by Eq.(\ref{57}). If we consider the scattering on a proton
we may choose $\mu$ to be around $1 \:{\rm GeV}$ but we do not
know the exact value. The same applies for $\Lambda$ which can be
taken to be around $200 \: {\rm MeV}$. However, changing
$\rho_c(0)$ by small amounts does not affect our results. We have
already discussed that the corresponding corrections for
$\lambda_s$ are parametrically very small. From another point of
view we first notice that Eq.(\ref{30}) has ``translational''
invariance in $Y$. Thus, changing $\rho_c(0)$, for example from $2
\log 5$ to $2 \log 3$ which describes equally good a hadronic
target, corresponds to a shift of the rapidity by a unit or so.
Recalling our results from Section 6, where we found that
$\lambda_s$ is effectively constant in a wide regions of $Y$
values, it is clear the error is small. On the contrary, the value
of $Q_s$ is sensitive to the initial conditions assumed.
\vspace{6pt}

\noindent {\bf (ii)} In Section 3 we have expanded our basic
$\rho_c$ and $\gamma_c$ equations, (\ref{26}) and (\ref{27}), to
first order in $\delta \gamma = \gamma_c - \gamma_0$. For the
Leading in $\omega$ kernel, we easily find that

\begin{equation}\label{76}
    \delta \gamma=
    \frac{2 \lambda_c (1-\gamma_0)
    [1 +b (1-\gamma_0) (1+\lambda_c)]}
    {(1-\gamma_0)^3 \chi''_0(\gamma_0) (1+\lambda_c)
    - 2 \lambda_c}=\frac{\lambda_c (1.85+0.591 \lambda_c)}
    {12 + 10 \lambda_c}.
\end{equation}

\noindent If we expand to second order in $\delta \gamma$ we
have to add on the right-hand side of Eq.(\ref{30}) the term

\begin{equation}\label{77}
    -\frac{(\delta \gamma)^2 \phi^{2,0}(\gamma_0,\lambda_c)}
    {2 b (1-\gamma_0)}.
\end{equation}

\noindent This term gives a very small contribution in
(\ref{30}). Numerically it is  $1-2 \%$ of the leading term
when we plug in the expected value of $\lambda_c$, the reason
being the large numerical value of $\chi_0''(\gamma_0)$ which
appears in the denominator in (\ref{76}).

\noindent What is more interesting, is that this kind of
correction can be neglected by making use of a more rigorous
argument. From the equations above we see that the contribution is
of order $\mathcal{O} (\lambda_c^2)$, which is of order
$\mathcal{O}(\omega^2)$ in view of (\ref{22}). Clearly this
overlaps with NNL in $\omega$ corrections. \vspace{6pt}

\noindent {\bf (iii)} In Section 5 we fixed the amplitude to be
constant on the saturation line. This will be exactly true in the
asymptotic region. In the intermediate region factors of the form
$\exp[\mathcal{O}(1/t^2)]\equiv \exp[g(t)]$ remain, because the
functions $h_1$ and $h_2$ are not equal to 1. These functions
though, turn out to be close to 1, and therefore it is harmless to
ignore these terms. Equivalently, we could multiply our solution
(\ref{43}) with $\exp[-g(t)]$, which gives an equally good
approximate solution since the remainder (\ref{44}) changes just
by a term $h_1(t) dg(t)/dt=\mathcal{O}(1/t^3)$.

\noindent There are also certain assumptions and approximations
that we made when deriving $\delta \xi_s$ and $\rho_s$ as given in
(\ref{49}) and (\ref{52}) respectively. We kept only the linear
term in $\delta \xi$ in the expansion of the Airy function in
(\ref{47}), we replaced $D$ by its asymptotic value $D_0$ and we
neglected $\xi^2/t$ and $\xi/t^2$ terms in the exponent of the
amplitude. If we relax these assumptions, we find that we should
add in (\ref{52}) terms of the form
$\mathcal{O}(\rho_c^{-2/3},\rho_c^{-1},...)$. Notice that there
are no $\mathcal{O}(\rho_c^{-1/3})$ terms. All these corrections
vanish when $\rho_c$ becomes large. Thus in (\ref{52}) all the
terms that survive in the asymptotic region, including the
constant one, have been kept. For completeness, we have checked
that if we include all the available terms up to order
$\mathcal{O}(\rho_c^{-1})$, then $\lambda_s$ is decreasing by
$\simeq 0.01$ in the intermediate $Y$ region. Therefore, we
wouldn't like to sacrifice the relative simplicity of (\ref{52})
and (\ref{58}) for such an unimportant error. \vspace{6pt}

\noindent {\bf (iv)} A crucial assumption that we made in order to
simplify the BFKL equation, was to drop the $\partial /\partial Y$
term that appeared inside the argument of the BFKL kernel in
(\ref{24}) (of course we did that after switching to the scaling
variable $\eta$; the argument cannot be applied directly in
(\ref{17})). One can understand that this term has a small effect,
since we are looking for a scaling solution. The dominant
$\partial /\partial Y$ term which appears on the left-hand side of
(\ref{24}) is already one of the weak terms of our equation and
therefore all other derivatives of this kind, which originate from
NL corrections will be even weaker. It is interesting to
understand this through a more rigorous argument. Let us consider
the most important contribution of such a term, which is obtained
by letting $\partial/\partial \eta \rightarrow \ -(1-\gamma_0)$
and expanding to 1st order in the 2nd argument in (\ref{24}). We
get

\begin{equation}\label{78}
    -\left(1-\frac{\eta}{\rho_c}\right)
    \frac{\phi^{0,1}(\gamma_0,\lambda_c)}{1-\gamma_0}
    \frac{\partial}{\partial Y}
    (T/\alpha_s)
\end{equation}

\noindent as a contribution to the left-hand side of our equation.
For the case of the leading in $\omega$ kernel, $\phi^{0,1}$ is

\begin{equation}\label{79}
    \phi^{0,1}(\gamma_0,\lambda_c)=
    -\frac{1}{(1-\gamma_0)(1+\lambda_c)^2}<0.
\end{equation}

\noindent We already see from (\ref{78}) that the correction is
suppressed by $1/\rho_c$, when compared to the dominant
$\partial/\partial Y$ term. Then we just follow the procedure of
Sections 3 and 4 and we find that $\psi(\xi,t)$ is still given by
(\ref{43}), where now the functions $h_1(t)$ and $h_2(t)$ are more
complicated but their asymptotic limit remains the same. At the
same time the $\rho_c$ equation changes because of the $\eta
\,\partial/\partial Y$ in (\ref{78}). This will add on the
right-hand side of Eq.(\ref{30}) a term

\begin{equation}\label{80}
    \lambda_c^2 \,
    \frac{2 \phi^{0,1}(\gamma_0,\lambda_c)}
    {\phi(\gamma_0,\lambda_c)-2 b \lambda_c}.
\end{equation}

\noindent However, as in case {\bf (ii)} this is of order
$\mathcal{O}(\omega^2)$ which overlaps with the NNL in $\omega$
corrections.\vspace{6pt}

\noindent{\bf (v)} We have considered three active flavors. A
fourth quark, the charm, cannot be treated as massless as required
by BFKL dynamics. Even if we set $N_f=4$, then $\lambda_s$ will
change by $\simeq 0.01$.

\section{Summary}

We have studied the energy dependence of the saturation momentum,
using the collinearly (RG) improved BFKL equation as our QCD
dynamics. A saturation boundary of constant amplitude for
dipole-hadron scattering has been used to describe the non-linear
effects in a simple but effective way. Following the evolution of
the system, we were able to determine this boundary as far as the
energy (or rapidity $Y$) is considered. The saturation momentum
$Q_s^2$ and its logarithmic derivative $\lambda_s$ depend on $Y$
in a complicated way, particularly at the Next to Leading Level.
However, we have found that $\lambda_s$ is decreasing very slowly,
so that it is practically constant for a wide region of $Y$ values
and therefore $Q_s^2$ is effectively described by the simple
exponential law $Q_s^2 \propto \Lambda^2 \exp (\lambda_s Y)$. In
the region of current phenomenological interest we found at the NL
level $\lambda_s=0.30 \div 0.29$ with very reliable accuracy. The
boundary (which imitates the non-linear effects), the RG
improvement of the BFKL kernel at Leading and NL level, and the
running of the coupling, were all important in obtaining
$\lambda_s(Y)$. The exact value of $Q_s$ cannot be determined in
our approach, because of a (not totally though) free
multiplicative constant. The amplitude exhibits geometrical
scaling in a certain region on the perturbative side of the
boundary, and its form is independent of the particular BFKL
dynamics considered.

\section*{Acknowledgments}

I wish to thank A.H.~Mueller for proposing the subject of this
work and for valuable discussions.

\appendix
\section*{Appendix}
\setcounter{section}{1}
\setcounter{equation}{0}
\numberwithin{equation}{section}

Here we construct the NL in $\omega$ kernel with an energy scale
choice $s_0=Q^2$ following Refs.\cite{Salam,CiaCol,CCS,Salam2}. We
start with the NL BFKL kernel \cite{FadLip,CamCia},

\begin{align}\label{A1}
    \chi_1(\gamma)=&-\frac{b}{2}
    [\chi_0^2(\gamma)+\chi'_0(\gamma)]
    -\frac{1}{4} \chi''_0(\gamma)
    \nonumber \\
    &-\frac{\pi^2 \cos(\pi \gamma)}
    {4 (1-2\gamma)\sin^2(\pi \gamma)}
    \bigg[
    3+\bigg(1+\frac{N_f}{N_c^3}\bigg)
    \frac{2+3 \gamma (1-\gamma)}
    {(3-2 \gamma) (1+2 \gamma)}
    \bigg]
    \nonumber\\
    &+\frac{3}{2}\zeta(3)
    +\bigg(\frac{67}{36}-\frac{\pi^2}{12}
    -\frac{5}{18}\frac{N_f}{N_c}\bigg)\chi_0(\gamma)
    +\frac{\pi^3}{4 \sin(\pi \gamma)}
    -\Phi(\gamma),
\end{align}

\noindent where

\begin{equation}\label{A2}
    \Phi(\gamma)=
    \frac{\pi^3}{6 \sin(\pi \gamma)}
    -\int_0^1 dx\,
    \frac{x^{-\gamma}+x^{\gamma-1}}{1+x}
    {\rm Li_2}(x),
\end{equation}

\noindent with the leading BFKL kernel $\chi_0(\gamma)$ and $b$ as
given in Eqs.(\ref{2}) and (\ref{5}) in the text. The pole
structure is different that the one given in (\ref{4}) because now
we will include the $N_f$ dependence. We need first to give the
splitting functions and their Mellin transforms, which are

\begin{equation}\label{A3}
    P_{gg}(z)=\frac{z}{(1+z)_+}
    +\frac{1-z}{z}+z(1-z)
    +b\,\delta(1-z),
\end{equation}

\begin{equation}\label{A4}
    P_{qg}=\frac{N_f}{2N_c}[z^2+(1-z)^2],
\end{equation}

\begin{align}\label{A5}
    A_{gg}(\omega)&=
    \int_0^1 dz\, z^{\omega} P_{gg}(\omega)
    \; -\frac{1}{\omega}
    \nonumber\\
    &=b-\frac{1}{1+\omega}+
    \frac{1}{2+\omega}-
    \frac{1}{3+\omega}
    -[\psi(2+\omega)-\psi(1)],
\end{align}

\begin{align}\label{A6}
    A_{qg}(\omega)&=
    \int_0^1 dz\, z^{\omega} P_{qg}(\omega)
    \nonumber\\
    &=\frac{N_f}{2N_c}
    \bigg(\frac{1}{1+\omega}+-
    \frac{2}{2+\omega}+
    \frac{2}{3+\omega}
    \bigg).
\end{align}

\noindent Define the function $A_{T}(\omega)$ by

\begin{equation}\label{A7}
    A_{T}(\omega)= A_{gg}(\omega)
    +\frac{C_F}{N_c} A_{qg}(\omega),
\end{equation}

\noindent with $C_F=(N_c^2-1)/2N_c$. We can write the pole
structure of $\chi_1(\gamma)$ as

\begin{align}\label{A8}
    \chi_1(\gamma)=&
    -\frac{1}{2 \gamma^{3}}
    -\frac{1}{2 (1- \gamma)^{3}}
    +\frac{A_{T}(0)}{\gamma^{2}}
    +\frac{A_{T}(0)-b}{(1-\gamma)^{2}}
    \nonumber \\
    &-F \bigg(\frac{1}{\gamma}+\frac{1}{1-\gamma}\bigg)
    +{\rm finite},
\end{align}

\noindent where $F$, which vanishes when $N_f=0$ causing the 
simple poles to go away, is

\begin{equation}\label{A9}
    F=\frac{N_f}{6N_c}
    \bigg(\frac{5}{3}+\frac{13}{6N_c^2}\bigg).
\end{equation}

\noindent So far this refers to a symmetric energy scale
$s_0=Q\mu$ and in order to transform $\chi_1(\gamma)$ to the scale
$s_0=Q^2$ we need to add the piece
$-\chi_0(\gamma)\chi'_0(\gamma_0)/2$ \cite{FadLip,CamCia}. This
will remove the triple pole at $\gamma=0$ and will double the one
at $\gamma=1$.

Now, our leading in $\omega$ kernel $\chi_0(\gamma,\omega)$ as
given in (\ref{15}) contains NL corrections which must be
subtracted from our NL kernel to avoid double counting. Iterating
$\omega=\bar{\alpha}_s \chi_0(\gamma,\omega)$ once, we easily find
that we must add to the NL kernel the term
$\chi_0(\gamma)/(1-\gamma)^2$. This will cancel the remaining
triple pole at $\gamma=1$.

There are still double and single poles\footnote{There is a small
ambiguity with the single poles emerging from the $N_f$ part of
the kernel \cite{CCS}. However, it is well-known that the
fermionic part is very small, and this ambiguity is essentially
unimportant for our purposes.}. Putting all the previous pieces
together and subtracting these remaining poles we find

\begin{align}\label{A10}
    \tilde{\chi}_1(\gamma)=&
    \chi_1(\gamma)
    -\frac{1}{2}\chi_0(\gamma)\chi'_0(\gamma)
    +\frac{\chi_0(\gamma)}{(1-\gamma)^2}
    \nonumber\\
    &-\frac{1}{\gamma}
    +F \bigg(\frac{1}{\gamma}+\frac{1}{1-\gamma} \bigg)
    -\frac{A_T(0)}{\gamma^2}
    -\frac{A_T(0)-b}{(1-\gamma)^2},
\end{align}

\noindent a function with no poles at all at $\gamma=0$ and
$\gamma=1$. Of course we need now to add the correct form of all
these subtracted terms that appear in the second line of the
previous equation. We obtain

\begin{align}\label{A12}
    \chi_1(\gamma,\omega)=&
    \tilde{\chi}_1(\gamma)
    +\frac{1}{\gamma}
    -F \bigg(\frac{1}{\gamma}+\frac{1}
    {1-\gamma+\omega} \bigg)
    \nonumber\\
    &+\frac{A_T(\omega)}{\gamma^2}
    +\frac{A_T(\omega)-b}{(1-\gamma+\omega)^2}.
\end{align}

\noindent Finally the full kernel is given as in Eq.(\ref{16}).

\end{document}